\documentclass[%
 aps,
prd,
twocolumn,
showpacs,superscriptaddress,nofootinbib,floatfix]{revtex4-2}

\usepackage{graphicx}
\usepackage{dcolumn}
\usepackage{bm}
\usepackage[mathlines]{lineno}

\usepackage{amsmath}   
\usepackage{graphicx}  
\usepackage{xspace}
\usepackage{units}
\usepackage{comment}
\usepackage{xcolor}
\usepackage{setspace}
\usepackage{multirow}

\usepackage{natbib}
\usepackage{hyperref}
\hypersetup{breaklinks=true}
\newcommand{\minerva}{MINER$\nu$A\xspace}

\newcommand{\numu}{\ensuremath{\nu_{\mu}}\xspace}
\newcommand{\numubar}{\ensuremath{\bar{\nu}_{\mu}}\xspace}

\newcommand{\gf}{\textsc{Geant4}\xspace}

\newcommand{\sizecheck}{0} 
\newcommand{\PRDsupp}{1}   
\ifnum\PRDsupp=0
  
\else
  
\fi

\begin{document}
\title{Improved constraint on the MINERvA medium energy neutrino flux using $\bar{\nu}e^{-} \!\rightarrow \bar{\nu}e^{-}$ data} 

\newcommand{\Rutgers}{Rutgers, The State University of New Jersey, Piscataway, New Jersey 08854, USA}
\newcommand{\Hampton}{Hampton University, Dept. of Physics, Hampton, VA 23668, USA}
\newcommand{\Dortmund}{Institute of Physics, Dortmund University, 44221, Germany }
\newcommand{\Otterbein}{Department of Physics, Otterbein University, 1 South Grove Street, Westerville, OH, 43081 USA}
\newcommand{\JMU}{James Madison University, Harrisonburg, Virginia 22807, USA}
\newcommand{\Florida}{University of Florida, Department of Physics, Gainesville, FL 32611}
\newcommand{\UCIrvine}{Department of Physics and Astronomy, University of California, Irvine, Irvine, California 92697-4575, USA}
\newcommand{\CBPF}{Centro Brasileiro de Pesquisas F\'{i}sicas, Rua Dr. Xavier Sigaud 150, Urca, Rio de Janeiro, Rio de Janeiro, 22290-180, Brazil}
\newcommand{\PUCP}{Secci\'{o}n F\'{i}sica, Departamento de Ciencias, Pontificia Universidad Cat\'{o}lica del Per\'{u}, Apartado 1761, Lima, Per\'{u}}
\newcommand{\INRM}{Institute for Nuclear Research of the Russian Academy of Sciences, 117312 Moscow, Russia}
\newcommand{\Jlab}{Jefferson Lab, 12000 Jefferson Avenue, Newport News, VA 23606, USA}
\newcommand{\Pittsburgh}{Department of Physics and Astronomy, University of Pittsburgh, Pittsburgh, Pennsylvania 15260, USA}
\newcommand{\Guanajuato}{Campus Le\'{o}n y Campus Guanajuato, Universidad de Guanajuato, Lascurain de Retana No. 5, Colonia Centro, Guanajuato 36000, Guanajuato M\'{e}xico.}
\newcommand{\Athens}{Department of Physics, University of Athens, GR-15771 Athens, Greece}
\newcommand{\Tufts}{Physics Department, Tufts University, Medford, Massachusetts 02155, USA}
\newcommand{\WM}{Department of Physics, William \& Mary, Williamsburg, Virginia 23187, USA}
\newcommand{\FNAL}{Fermi National Accelerator Laboratory, Batavia, Illinois 60510, USA}
\newcommand{\Purdue}{Department of Chemistry and Physics, Purdue University Calumet, Hammond, Indiana 46323, USA}
\newcommand{\MCLA}{Massachusetts College of Liberal Arts, 375 Church Street, North Adams, MA 01247}
\newcommand{\UMD}{Department of Physics, University of Minnesota -- Duluth, Duluth, Minnesota 55812, USA}
\newcommand{\Northwestern}{Northwestern University, Evanston, Illinois 60208}
\newcommand{\UNI}{Facultad de Ciencias, Universidad Nacional de Ingenier\'{i}a, Apartado 31139, Lima, Per\'{u}}
\newcommand{\Rochester}{Department of Physics and Astronomy, University of Rochester, Rochester, New York 14627 USA}
\newcommand{\Austin}{Department of Physics, University of Texas, 1 University Station, Austin, Texas 78712, USA}
\newcommand{\USM}{Departamento de F\'{i}sica, Universidad T\'{e}cnica Federico Santa Mar\'{i}a, Avenida Espa\~{n}a 1680 Casilla 110-V, Valpara\'{i}so, Chile}
\newcommand{\Geneva}{University of Geneva, 1211 Geneva 4, Switzerland}
\newcommand{\Chicago}{Enrico Fermi Institute, University of Chicago, Chicago, IL 60637 USA}
\newcommand{\hired}{}
\newcommand{\OregonState}{Department of Physics, Oregon State University, Corvallis, Oregon 97331, USA}
\newcommand{\oxford}{Oxford University, Department of Physics, Oxford, OX1 3PJ United Kingdom}
\newcommand{\umiss}{University of Mississippi, Oxford, Mississippi 38677, USA}
\newcommand{\upenn}{Department of Physics and Astronomy, University of Pennsylvania, Philadelphia, PA 19104}
\newcommand{\AMU}{Department of Physics, Aligarh Muslim University, Aligarh, Uttar Pradesh 202002, India}
\newcommand{\wroclaw}{University of Wroclaw, plac Uniwersytecki 1, 50-137 Wroa\l{}aw, Poland}
\newcommand{\Mohali}{Department of Physical Sciences, IISER Mohali, Knowledge City, SAS Nagar, Mohali - 140306, Punjab, India}
\newcommand{\CINVESTAV}{Departamento de Fisica Col. San Pedro Zacatenco, 07360 Mexico, DF, Av. Instituto PolitÃ©cnico Nacional, Mexico}
\newcommand{\york}{York University, Department of Physics and Astronomy, Toronto, Ontario, M3J 1P3 Canada}
\newcommand{\ND}{Department of Physics, University of Notre Dame, Notre Dame, Indiana 46556, USA}
\newcommand{\ICL}{The Blackett Laboratory,  Imperial College London,  London SW7 2BW, United Kingdom}
\newcommand{\warwick}{Department of Physics, University of Warwick, Coventry, CV4 7AL, UK}

\newcommand{\mascencioThanks}{Now at Iowa State University, Ames, IA 50011, USA}
\newcommand{\amitbashyalThanks}{Now at  High Energy Physics/Center for Computational Excellence Department, Argonne National Lab, 9700 S Cass Ave, Lemont, IL 60439}
\newcommand{\finerThanks}{Now at Los Alamos National Laboratory, Los Alamos, New Mexico 87545, USA}
\newcommand{\kleykampThanks}{now at Department of Physics and Astronomy, University of Mississippi, Oxford, MS 38677}
\newcommand{\emilymaherThanks}{Department of Physics}
\newcommand{\bamThanks}{Now at University of Minnesota, Minneapolis, Minnesota 55455, USA}
\newcommand{\byaeggyThanks}{Now at Department of Physics, University of Cincinnati,  Cincinnati, Ohio 45221, USA}


\author{L.~Zazueta}                       \affiliation{\WM}
\author{S.~Akhter}                        \affiliation{\AMU}
\author{Z.~~Ahmad~Dar}                    \affiliation{\WM}  \affiliation{\AMU}
\author{F.~Akbar}                         \affiliation{\AMU}
\author{V.~Ansari}                        \affiliation{\AMU}
\author{M.~V.~Ascencio}\thanks{\mascencioThanks}  \affiliation{\PUCP}
\author{M.~Sajjad~Athar}                  \affiliation{\AMU}
\author{A.~Bashyal}\thanks{\amitbashyalThanks}  \affiliation{\OregonState}
\author{A.~Bercellie}                     \affiliation{\Rochester}
\author{M.~Betancourt}                    \affiliation{\FNAL}
\author{J.~L.~Bonilla}                    \affiliation{\Guanajuato}
\author{A.~Bravar}                        \affiliation{\Geneva}
\author{T.~Cai}                           \affiliation{\Rochester}
\author{G.A.~D\'{i}az~}                   \affiliation{\Rochester}
\author{H.~da~Motta}                      \affiliation{\CBPF}
\author{J.~Felix}                         \affiliation{\Guanajuato}
\author{L.~Fields}                        \affiliation{\ND}
\author{A.~Filkins}                       \affiliation{\WM}
\author{R.~Fine}\thanks{\finerThanks}     \affiliation{\Rochester}
\author{A.M.~Gago}                        \affiliation{\PUCP}
\author{H.~Gallagher}                     \affiliation{\Tufts}
\author{A.~Ghosh}                         \affiliation{\USM}  \affiliation{\CBPF}
\author{R.~Gran}                          \affiliation{\UMD}
\author{E.Granados}                       \affiliation{\Guanajuato}
\author{D.A.~Harris}                      \affiliation{\york}  \affiliation{\FNAL}
\author{S.~Henry}                         \affiliation{\Rochester}
\author{D.~Jena}                          \affiliation{\FNAL}
\author{S.~Jena}                          \affiliation{\Mohali}
\author{J.~Kleykamp}\thanks{\kleykampThanks}  \affiliation{\Rochester}
\author{A.~Klustov\'{a}}                  \affiliation{\ICL}
\author{M.~Kordosky}                      \affiliation{\WM}
\author{D.~Last}                          \affiliation{\upenn}
\author{A.~Lozano}                        \affiliation{\CBPF}
\author{X.-G.~Lu}                         \affiliation{\warwick}  \affiliation{\oxford}
\author{E.~Maher}\thanks{\emilymaherThanks}  \affiliation{\MCLA}
\author{S.~Manly}                         \affiliation{\Rochester}
\author{W.A.~Mann}                        \affiliation{\Tufts}
\author{K.S.~McFarland}                   \affiliation{\Rochester}
\author{B.~Messerly}\thanks{\bamThanks}   \affiliation{\Pittsburgh}
\author{J.~Miller}                        \affiliation{\USM}
\author{O.~Moreno}                        \affiliation{\WM}  \affiliation{\Guanajuato}
\author{J.G.~Morf\'{i}n}                  \affiliation{\FNAL}
\author{J.K.~Nelson}                      \affiliation{\WM}
\author{C.~Nguyen}                        \affiliation{\Florida}
\author{A.~Olivier}                       \affiliation{\Rochester}
\author{V.~Paolone}                       \affiliation{\Pittsburgh}
\author{G.N.~Perdue}                      \affiliation{\FNAL}  \affiliation{\Rochester}
\author{K.-J.~Plows}                      \affiliation{\oxford}
\author{M.A.~Ram\'{i}rez}                 \affiliation{\upenn}  \affiliation{\Guanajuato}
\author{D.~Ruterbories}                   \affiliation{\Rochester}
\author{H.~Schellman}                     \affiliation{\OregonState}
\author{C.J.~Solano~Salinas}              \affiliation{\UNI}
\author{H.~Su}                            \affiliation{\Pittsburgh}
\author{M.~Sultana}                       \affiliation{\Rochester}
\author{E.~Valencia}                      \affiliation{\WM}  \affiliation{\Guanajuato}
\author{N.H.~Vaughan}                     \affiliation{\OregonState}
\author{A.V.~Waldron}                     \affiliation{\ICL}
\author{B.~Yaeggy}\thanks{\byaeggyThanks}  \affiliation{\USM}

\collaboration{The MINER$\nu$A Collaboration}\ \noaffiliation

\date{\today}

\begin{abstract}
Processes with precisely known cross sections, like neutrino electron elastic scattering ($\nu e^{-} \!\rightarrow \nu e^{-}$) and inverse muon decay ($\nu_\mu e^{-} \!\rightarrow \mu^{-} \nu_e$) have been used by \minerva to constrain the uncertainty on the NuMI neutrino beam flux. This work presents a new measurement of neutrino elastic scattering with electrons using the medium energy \numubar enhanced NuMI beam. A sample of 578 events after background subtraction is used in combination with the previous measurement on the \numu beam and the inverse muon decay measurement to reduce the uncertainty on the \numu flux in the \numu-enhanced beam from 7.6\% to 3.3\% and the \numubar flux in the \numubar-enhanced beam from 7.8\% to 4.7\%.
\end{abstract}

\ifnum\sizecheck=0  
\maketitle
\fi

\section{Introduction}
The neutrino beam is a critical component of accelerator-based neutrino oscillation experiments. Produced by colliding a high energy proton beam into a stationary target and then focusing the produced charged hadrons with one or more magnetic horns, neutrino beams provide an intense source of neutrinos with a tunable neutrino energy. Neutrino beams are used by current experiments such as T2K \cite{PhysRevD.103.112008}, NO$\nu$A \cite{Nova2019PhysRevLett.123.151803} and MicroBooNE \cite{microbooneccqe-PhysRevLett.125.201803}, and in future oscillation experiments such as DUNE \cite{Acciarri:2015uup}, SBN \cite{SBN} and T2HK \cite{Abe:2018uyc}. 

Neutrino beams carry large uncertainties on the total number of neutrinos produced and their energy spectrum. This is due to the underlying uncertainty on the hadron production multiplicity and kinematics, as well as to the uncertainty in the parameters related to the focusing components. Oscillation experiments can deal with this by using near detector measurements and adjusting their \textit{a priori} model. This procedure correlates the flux and the cross section, which also has significant uncertainty.

The neutrino flux model can be improved by using external hadron production data, reducing the flux error to around 8\% \cite{Aliaga:2016oaz,T2K:2020txr}. Another way to improve the neutrino flux prediction is by measuring processes with cross sections with known energy dependence.
\minerva has previously used events with low hadronic recoil to constrain the energy dependence of the flux \cite{Amitfluxfit:2021mpk}. 

Neutrino electron elastic scattering provides another known cross section, predicted by the Standard Model since it is a purely leptonic process. The \minerva collaboration has demonstrated that this process can be used to reduce the uncertainty on the flux using data taken during the NuMI low energy (LE) beam period \cite{Park:2015eqa}, and during the medium energy (ME) neutrino-enhanced beam \cite{PhysRevD.100.092001} (referred to as \numu-mode). Additionally, \minerva has explored the use of another purely leptonic process, inverse muon decay (IMD), to constrain the high energy region of the flux \cite{DanIMD}. 

This paper reports the measurement of the final state electron energy distribution for neutrino electron elastic scattering interactions observed in \minerva, after background subtraction and efficiency correction. The data were taken using the NuMI ME antineutrino enhanced beam (\numubar-mode). This measurement is used in combination with the \numu-mode neutrino-electron elastic scattering and the inverse muon decay results to produce an improved flux constraint that can be applied to \minerva cross section measurements. This work illustrates the procedure that can be followed in other accelerator-based neutrino experiments.

Section \ref{sec:numi} describes the NuMI beam line and its simulation. Section \ref{sec:minerva} describes the \minerva detector and the  simulation used. The event reconstruction and selection is described in Section \ref{sec:event_selection}. The procedure used to constrain and subtract the background is described in section \ref{sec:background}. The resulting uncertainties on the number of neutrino-electron elastic scattering events are discussed in Section \ref{sec:systematics}. The procedure and result from using the combined measurements to constrain the neutrino flux are described in Section \ref{sec:flux} and the conclusion is presented in Section \ref{sec:conclusion}.

\section{NuMI Beam line and Simulation}
\label{sec:numi}
The NuMI beam \cite{Adamson:2015dkw} is the neutrino source of the \minerva experiment. It starts with a 120 GeV proton beam hitting a carbon target, producing a hadronic shower focused in the forward direction by two magnetic horns. The beam is aimed 58 mrad downward through a 675 meter decay pipe in which the secondary mesons decay into neutrinos. NuMI spills are delivered in 6 bunches in a window of 10 $\mu$s. The polarity of the magnetic horns sets the predominant helicity of the beam. The forward horn current (FHC) polarity produces predominantly muon neutrinos. The reverse horn current (RHC) polarity produces predominantly muon antineutrinos. The neutrino flux prediction used by \minerva is derived from a \gf simulation of the NuMI beamline. The simulation is reweighted to agree with external proton-on-carbon hadron production data in a flux-tuning procedure developed by \minerva \cite{Aliaga:2016oaz}. This modified neutrino flux prediction is the \textit{a priori} constraint used in section \ref{sec:flux}. 
This analysis uses data taken in a period between June 2016 and February 2019 during the \numubar-mode and corresponds to an integrated $1.2\times 10^{21}$ protons on target (POT).

\section{MINERVA Experiment and Simulation}
\label{sec:minerva}
\minerva (described in detail in Ref.\cite{Aliaga:2013uqz}) consists of 120 hexagonal active tracking modules, each made of two 1.7 cm thick planes, built of triangular scintillator strips. The strips in each plane are arranged in three different orientations or views: 0$^\circ$ and $\pm$60$^\circ$ from the vertical, allowing fine-grained, three-dimensional track reconstruction. The tracker region consists of 62 tracking modules. Electromagnetic calorimetry is accomplished on the side of the detector by a lead collar between each scintillator plane covering the outer 15 cm of each plane, and downstream of the tracker region with lead plates on the next 10 modules covering the full hexagonal plane. Hadronic calorimetry is performed by placing steel planes between scintillator planes on the last 20 most downstream modules. These detector regions are referred to as the electromagnetic (Ecal) and hadronic (Hcal) calorimeters respectively, and provide full containment for forward going electromagnetic showers which are a signature of elastically scattered electrons. Upstream of the tracker region are the nuclear targets which serve as passive targets for cross section measurements. Events with activity near the nuclear targets are not used in this analysis. 

Scintillator light from the strips is collected by wavelength-shifting fibers and directed to photomultiplier tubes. The output signal is read out using the data acquisition system described in Ref. \cite{Perdue:2012hg}. The energy scale is calibrated by a sample of muons produced by the interaction of the beam with the rocks upstream of the cavern that houses the \minerva detector, and is further cross checked using electrons coming from muon decays and a sample of $\pi^0 \!\rightarrow \gamma \gamma$. The time resolution of individual hits is better than 4 ns.

Neutrino interactions at \minerva are simulated using the GENIE neutrino event generator version 2.12.6 \cite{Andreopoulos:2009rq}\cite{Andreopoulos:2015wxa}. The quasi-elastic neutrino-nucleus interactions are simulated with a relativistic Fermi gas model \cite{SMITH1972605} and the Llewellyn-Smith formalism \cite{LLEWELLYNSMITH1972261}. The quasi-elastic model is modified with a weak charge screening correction (Random phase approximation or RPA) \cite{Nieves:2004wx}. The procedure is described in Ref. \cite{Gran:2017psn}. Another modification is the addition of the interaction mode in which the neutrino scatters off a correlated pair of nucleons, leaving two holes in the nucleus (2p2h). The Valencia model is used to simulate this interaction mode \cite{PhysRevC.83.045501}. The Rein-Sehgal models have been implemented for resonance \cite{REIN198179} and coherent pion production \cite{REIN198329}. \minerva additionally includes a simulation of diffractive neutral-current $\pi^0$ production off hydrogen based on Rein's model \cite{Rein:1986cd} that is available in GENIE. Deep inelastic scattering is simulated using the Bodek-Yang model \cite{Bodek_2003}. Intranuclear rescattering is simulated using the GENIE INTRANUKE-hA package.

Propagation of particles through the \minerva detector is modeled with a simulation based on \gf version 4.9.4.p02 with the QGSP BERT physics list. Activity from overlapping events and dead time is simulated by overlaying data beam spills on top of simulated events. 

\section{Event Reconstruction and Selection}
\label{sec:event_selection}
Elastic neutrino-electron scattering interactions are observed in \minerva as very forward electromagnetic showers. The scattered electron travels through the detector as a minimum ionizing particle until it starts an electromagnetic shower. As a result, the track starts thin and widens as it propagates, creating a cone-shaped track until it stops after depositing all its kinetic energy in the detector. It is not possible to distinguish the flavor of incident neutrino or antineutrino from the final state electron. The signal definition includes events coming from both muon and electron flavor, for both neutrinos and antineutrinos. Simulation predicts that the final selection is comprised of about 70\% \numubar, 20\% $\bar{\nu}_e$ and 10\% \numu and $\nu_e$. The high number of $\bar\nu_e$ events is due to the larger cross section of the $\nu_e$-nucleus background, which is three orders of magnitude larger than the neutrino-electron cross section. 

\begin{figure}[t]
\centering
  \includegraphics[width=0.48\textwidth]{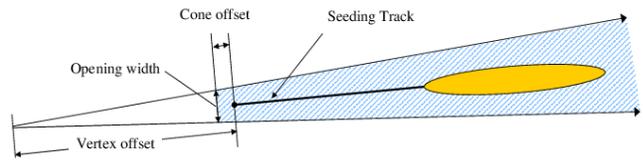}
\caption{An illustration of the cone algorithm used to search for hits that belong to the electromagnetic shower. The cone offset is 50 mm, the opening width is 80 mm, and cone opening angle is $10^\circ$. }
\label{fig:cone}
\end{figure}

Neutrinos reach \minerva during the 10 $\mu$s NuMI spill. The energy and time information of all hits is  recorded and later grouped in time forming ``time slices". Later spatial information is used to form clusters of hits. A Kalman filter is used to estimate the location of the vertex and angle of the track left by the electron. Sometimes the track starts to shower early, resulting in a short track, and a chi-square fit is used to assign a vertex and angle. The vertex is used as a seed for a cone algorithm that adds up all the energy inside the cone. The cone has an opening angle of $10^\circ$ and is placed such that the width is 80 mm at a point 50 mm upstream of the vertex. A sketch is shown in Fig. \ref{fig:cone}.

For this analysis, events are selected if the reconstructed vertex is within the hexagon with an apothem of 88.125 cm and within the 112 central planes of the tracker. This amounts to a fiducial mass of 5.99 metric tons. The upstream region is not used in the selection. The cone is extended downstream until it cannot find more hits. The energy of the hits inside the cone is added up taking into account the different calorimetry of the passive materials on the Ecal and Hcal. The energy (angle) resolution is 60 MeV ($0.7^\circ$) in the lowest bin of the electron energy spectrum at 0.8-2 GeV, and 40 MeV ($0.3^\circ$) in the highest energy bin covering electrons of energy $>$9 GeV.

The selection cuts are the same as used in the previous $\nu_\mu e$ elastic analysis by \minerva \cite{Park:2015eqa}, \cite{PhysRevD.100.092001}. The selection cuts were chosen using simulation to maximise efficiency and to minimise background.

Events are required to have a minimum total energy ($E_e$) of 0.8 GeV to assure good quality in angle and energy reconstruction and increase the sample purity by rejecting a large fraction of \numu neutral current events. Events coming from \numu CC would have a muon in the final state that would reach the edge of the detector. These events are removed by rejecting events that reach the sides and back of the detector, where the Ecal would have stopped an electromagnetic shower. 

\begin{figure}[t]
\centering
  \includegraphics[width=0.48\textwidth]{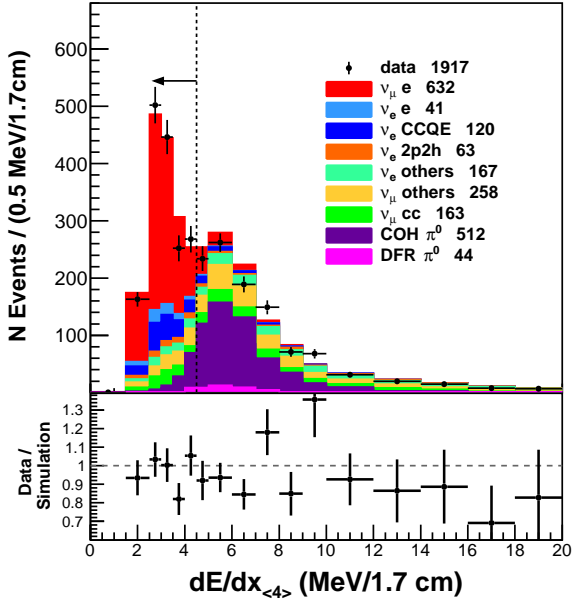}
\caption{Average energy deposition in the first four planes of the electron candidate track for events passing all other cuts after background tune (above) and the ratio of data to simulation (below).  The error bars on the data points include statistical uncertainties only.  The error bars on the ratio include both statistical uncertainties in data and statistical and systematic uncertainties in the simulation. Backgrounds have been tuned using the procedure described in Sec.~\ref{sec:background}. The dotted line and arrow indicated the selected sample.}
\label{fig:dedx}
\end{figure}

\begin{figure}[b]
\centering
\includegraphics[width=0.48\textwidth]{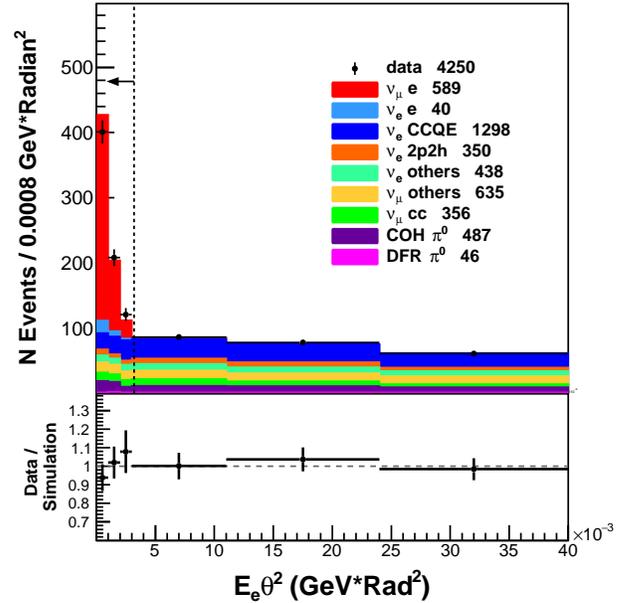} 
\caption{Product of the energy of the electron candidate and the square of its scattering angle with respect to the beam direction after background tune (above) and ratio of data to simulation (below). The error bars on the data include statistical uncertainties only. The error bars on the ratio include both statistical uncertainties in data and statistical and systematic uncertainties in the simulation. The dotted line and arrow indicates the selected sample. Backgrounds have been tuned using the procedure described in Sec.~\ref{sec:background}. The cut in $Q^2_{ QE}$ is not applied here to better show the sideband region.}
\label{fig:eth2}
\end{figure}

To remove hadronic background, at least 80\% of the total energy deposited in the Ecal and Hcal has to be deposited in the Ecal. Additionally, tracks are required to bend less than 9 degrees to reject tracks with overlapping hadron scattering. \minerva views are used to discriminate between overlapping hadron track and the electromagnetic showers. Since the plane views follow a XUXV pattern, it is expected that a shower would deposit 50\% of its energy on the X-view and 25\% on each of the U and V-views. To select events of this nature the energy ratios $E_{XUV}$ and $E_{UV}$ are formed, as follows:
\begin{align}
    E_{XUV} = & \frac{E_X-E_U-E_V}{E_X+E_U+E_V} \\
    E_{UV} = & \frac{E_U - E_V}{E_U+E_V}.
\end{align}
where $E_{X(U,V)}$ is the energy deposited in the X (U,V) view. Electron candidates are required to satisfy $E_{XUV}<0.28$ and $|E_{UV}|<0.5$.

The highest energy-weighted RMS distance transverse to the center of the shower between the three views must be less than 60 mm, and less than 20 mm in the first third of the shower length. Also, the energy within 5 cm of the outside boundary of the shower cone is required to be less than 120 MeV for events with less than 7 GeV of reconstructed energy. Otherwise, energy in this region must be less than $7.8E_e+65$ MeV to improve the purity for high energy events.

To remove the photon background from decaying $\pi^0$ mesons a cut is applied to the energy deposition per unit length in the first four detector planes of the track ($dE/dx_{\langle 4 \rangle}$). When a photon starts a shower, it begins by producing a $e^- e^+$ pair. Such showers would have about the double of $dE/dx$ of a shower initiated by an electron. This provides a way to separate photon-like and electron-like showers. Events are required to have $dE/dx_{\langle 4 \rangle}<4.5$ MeV/1.7cm. Figure \ref{fig:dedx} shows the $dE/dx_{\langle 4 \rangle}$ distribution with all other cuts applied. Additionally, to remove $\pi^0$ events where one or two photons could propagate a distance away from the true interaction vertex before converting, events with 300 MeV in a 30 cm-diameter cylinder projected upstream of the reconstructed vertex are rejected. Events are checked to be consistent with single shower on the transverse direction by looking for two peaks in the energy deposited in the Ecal, and in the longitudinal direction by checking that the distant from the start of the shower to the plane with the maximum energy deposition is consistent with an electromagnetic shower propagation in scintillator.

\begin{figure}[t]
\centering
\includegraphics[width=0.48\textwidth]{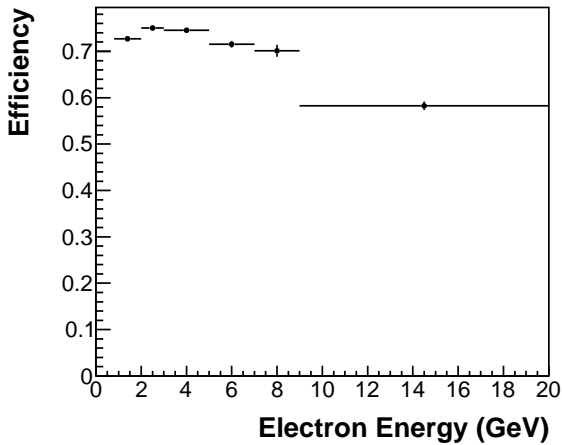} 
\caption{Efficiency of neutrino-electron scattering candidates after all selection cuts.}
\label{fig:efficiency}
\end{figure}

The remaining source of background comes from charged-current quasi-elastic (CCQE) interactions from $\bar\nu_e$, that is, $\bar\nu_e p \rightarrow e^{+}n$. The most effective cut to isolate the signal comes from the constrained kinematics of neutrino-electron elastic scattering, which obeys 
\begin{equation}
    E_e\theta^2<2m_e,
\end{equation}
where $E_e$ is the electron candidate energy and $\theta$ is its scattering angle in radians with respect of the beam direction.
The distribution for this quantity is shown in Fig. \ref{fig:eth2}. Events are required to have $E_e\theta^2<0.0032$ GeV rad$^2$. To remove any remaining high energy $\nu_e$ CCQE events that pass the kinematic cut, a cut is applied on the four-momentum transfer, which is calculated under the assumption of quasi-elastic kinematics. The reconstructed neutrino energy $E_{QE}$, and the squared four-momentum transfer $Q^2_{ QE}$, are
\begin{align}
    E_{\nu}^{ QE} = & \frac{m_p E_e-m_{e}^2/2}{m_p - E_e + p_e \cos\theta}\\
    Q^{2}_{ QE} & = 2m_p \left( E_{\nu}^{QE} - E_e \right),
\end{align}
where $m_p$ is the mass of the proton, $p_e$ is the electron momentum. Event candidates are required to have $Q^2_{ QE}$ less than 0.02 GeV$^2$.

The selection efficiency of signal events after all cuts is shown in Fig.~\ref{fig:efficiency}.

\section{Background Subtraction}
\label{sec:background}
After all the cuts are applied, the selected sample consist of 898 neutrino-electron elastic scattering candidates. The simulation predicts 921 events, from which 601 are signal events and 320 are background events. Neutral current (NC) interactions from \numu amount to 38$\%$ of the background, concentrated between 0.8 and 2 GeV, with exception for neutral pions produced in $\nu_\mu$-nucleus coherent interactions which are also present at higher energies. Another 28$\%$ of the background comes from quasi-elastic events from $\nu_e$ with a forward going shower and a non-visible neutron in the final state. 

\begin{figure}[t]
\centering
\includegraphics[width=0.45\textwidth]{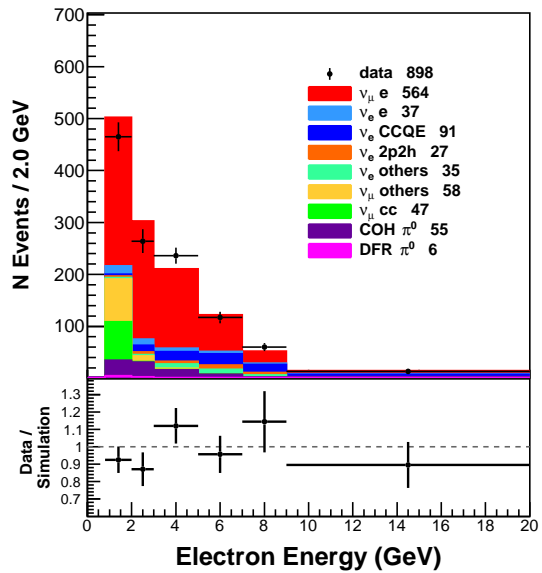} 
\caption{Reconstructed electron energy of the final sample in data and simulation after background tune (above) and the ratio of data to simulation (below). The error bars on the data  include statistical uncertainties only. The error bars on the ratio include both statistical uncertainties in data and statistical and systematic uncertainties in the simulation. Backgrounds have been tuned using the procedure described in Sec.~\ref{sec:background}. The highest energy bin includes all events with $E_e>9$ GeV, including events with $E_e>20$ GeV.}
\label{fig:energy}
\end{figure}

The background predicted by the GENIE simulation is constrained using four kinematic sidebands. The four sidebands are defined using the kinematic quantities $E_e\theta^2$ and $dE/dx_{\langle 4 \rangle}$. Sidebands 1-3 have $0.005<E_e\theta^2<0.112$ GeV rad$^2$ and $dE/dx_{\langle 4 \rangle}<20$ MeV/1.7cm. The cuts on $Q^2_{ QE}$ and the transverse energy spread of the first third of the shower are removed to improve statistics on the sidebands.

Sideband 1 requires events which the single plane minimum energy deposition between the second and sixth plane of the track $dE/dx_{\text{min}}$ is greater than 3 MeV. Sideband 2 and 3 have $dE/dx_{\text{min}} < 3$ MeV, and are further divided by requiring that reconstructed energy is $E_e <$ 1.2 GeV for sideband 2 and $E_e>1.2$ GeV for sideband 3. Sideband 4 is defined at the region of $dE/dx$ where the peak of the photon-like track is located. This sideband has all the same cuts that the signal region except that events must fall into 4.5 MeV/1.7 cm$ < dE/dx_{\langle 4 \rangle} <$ 10 MeV/1.7 cm.

The sidebands are designed to constrain three background categories: coherent neutral pion production, background from \numu (excluding coherent $\pi^0$), and background from $\nu_e$. The normalization of the \numu and $\nu_e$ background are allowed to float. Coherent neutral pion production is fitted to six bins of electron energy to better fit the photon-like peak on $dE/dx$ where events are originally under-predicted. This is also motivated by discrepancies with GENIE seen in Charged-current coherent pion production that vary with the pion's energy \cite{Mislivec:2017qfz}.

The resulting scale factors from the fit are shown in Table \ref{tab:backgrounds} along with the scale factors used in the \numu-mode analysis. The reconstructed electron energy distribution with the constrained background is shown in Fig. \ref{fig:energy}.

\begin{table}[b]
\centering
\caption{Scale factors from the fit to the background components on the kinematic sidebands. Uncertainties are statistical. For the \numubar-mode the normalization of \numu CC and \numu NC are set the same to avoid the strong anti correlation between them when calculating the fit. \numu-mode result from Ref. \cite{PhysRevD.100.092001}. }

\begin{tabular}{ c | c | c}
\hline \hline
Process &  \numubar-mode  &  \numu-mode \\ 
\hline
$\nu_e$ & $1.02\pm0.02$ & $0.87\pm0.03$ \\
$\nu_\mu$ CC & $0.93\pm0.03$ & $1.08\pm0.04$ \\
$\nu_\mu$ NC & $0.93\pm0.03$ & $0.86\pm0.04$ \\
NC COH $0.8 < E_e < 2.0$ GeV & $1.6\pm0.2$ & $0.9 \pm 0.2$\\
NC COH $2.0 < E_e < 3.0$ GeV & $2.1\pm0.3$ & $1.0 \pm 0.3$ \\
NC COH $3.0 < E_e < 5.0$ GeV & $1.8\pm0.2$ & $1.3 \pm 0.2$ \\
NC COH $5.0 < E_e < 7.0$ GeV & $2.1\pm0.4$ & $1.5 \pm 0.3$\\
NC COH $7.0 < E_e < 9.0$ GeV & $1.2\pm0.7$ & $1.7 \pm 0.8$ \\
NC COH $9.0 < E_e$ & $0.8\pm0.6$ & $3.0 \pm 0.9$ \\
\hline \hline
\end{tabular}
\label{tab:backgrounds}
\end{table}

\section{Systematic Uncertainties}
\label{sec:systematics}
\begin{figure}[t]
\centering
  \includegraphics[width=0.45\textwidth]{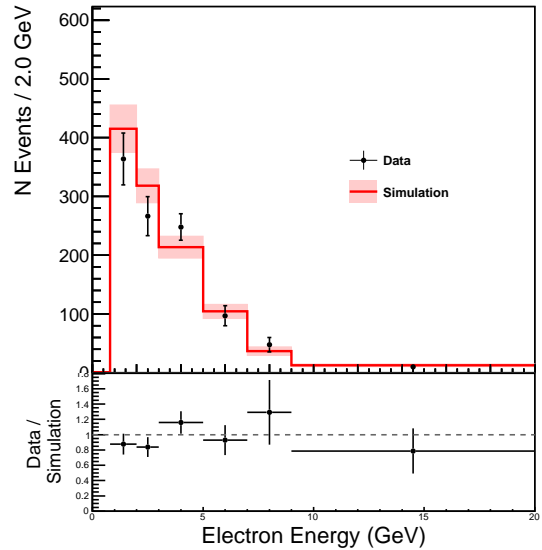} 
\caption{Reconstructed electron energy after background subtraction and efficiency correction in data and simulation (above) and the ratio of data to simulation (below).  The data error bars include both statistical and systematic uncertainties. The error bar in the simulation include statistical uncertainty as well as the systematic uncertainty coming from the flux. The highest energy bin includes all events with $E_e>9$ GeV, including events with $E_e>20$ GeV.}
\label{fig:effcor}
\end{figure}

\begin{figure}[t]
\centering
  \includegraphics[width=0.45\textwidth]{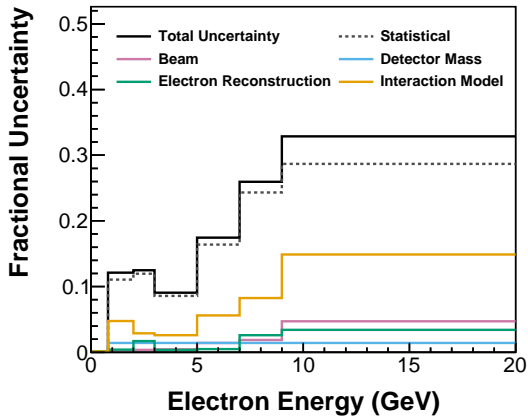} 
\caption{Summary of fractional systematic uncertainties on the the background subtracted, efficiency corrected, electron energy spectrum.}
\label{fig:uncertainty}
\end{figure}

The background-subtracted efficiency-corrected electron energy spectrum and its uncertainty are shown in Fig. \ref{fig:effcor} and Fig. \ref{fig:uncertainty}, respectively. The fractional uncertainty on the total number of neutrino electron elastic scattering events is shown in Tab. \ref{tab:error_summary} along with the uncertainties of the \numu-mode analysis for comparison. Uncertainties are evaluated by varying underlying parameters of a given model within their uncertainties. Each of these variations produces a new simulation prediction which is carried through selection, background subtraction, and efficiency correction. A covariance matrix of the electron energy spectrum is obtained for each variation. For models with more than one variation, an average of the covariance matrices is used for the error estimation.

Uncertainties are grouped in three categories: electron reconstruction, beam, and interaction model. The detector mass uncertainty is added as an uncertainty in the rate to facilitate the constraint procedure in Sec. \ref{sec:flux}.

\subsection{Electron Reconstruction Uncertainties}
The way that muon and electron tracks are seeded is the same. The tracking efficiency is estimated by projecting backward muon tracks that reach the MINOS near detector and comparing them with muon tracks in \minerva.  The difference between data and simulation is taken as a systematic uncertainty, and for the ME beam is 0.4\%. 

Uncertainty on the electromagnetic energy scale was studied by comparing the energy of reconstructed $\pi^0$ candidates in charged-current \numu events between data and simulation. The $\pi^0$ sample indicated a 5.8\% mismodeling of the energy scale of the electromagnetic calorimeter. The Ecal energy deposition is adjusted by 5.8\%, and an overall uncertainty in the electromagnetic response of 1.5\% is applied based on the precision of the $\pi^0$ sample. This results in a 0.20\% uncertainty on the total event rate.

\begin{table}[b]
\centering
\caption{Uncertainties on total number of neutrinos elastic scattering off electrons in \minerva after background subtraction and efficiency correction. Uncertainties from the \numu-mode analysis \cite{PhysRevD.100.092001} are shown for comparison.}
\begin{tabular}{ c | c | c }
\hline \hline
Source &  \multicolumn{2}{c}{Uncertainty (\%)} \\ 
 &  \numubar-mode  & \numu-mode \\
\hline
Beam & 0.22 & 0.21 \\
Electron Reconstruction & 0.20 & 0.57 \\
Interaction Model & 3.74 & 1.68 \\
Detector Mass & 1.40 & 1.40 \\ \hline
Total Systematic & 4.06 & 2.27 \\ \hline
Statistical & 5.49 & 4.17 \\ \hline
Total & 6.83 & 4.75 \\ 
\hline \hline
\end{tabular}
\label{tab:error_summary}
\end{table}

\subsection{Beam Uncertainties}
Uncertainties from the beam come in to the measurement from the background and efficiency correction. The main component comes in the form of uncertainties arising from the hadron production model and focusing elements such as the current of the focusing horns of the NuMI beam. These uncertainties are estimated using the procedure developed by \minerva for the LE configuration \cite{Aliaga:2016oaz}. Uncertainty on the neutrino-electron scattering rate due to uncertainties in the flux model is 0.2\%. 
The uncertainty in the beam angle is estimated by looking at the angular spectra of muons from charged-current \numu candidates with low hadron recoil in data and simulation. The uncertainty in the beam angle is 0.5 mrad, which gives an uncertainty in the neutrino-electron elastic scattering rate of 0.09\%.

\subsection{Interaction Model Uncertainties}
The biggest source of systematic uncertainty comes from model uncertainties and enter the measurement through the predicted background. The interaction model uncertainties for this \numubar analysis are higher than those for the corresponding \numu-mode analysis \cite{PhysRevD.100.092001} because the number of background events is higher. This is because background with one electromagnetic shower and final state neutrons are more difficult to reject, and final state neutrons are more common in \numubar-mode (for example: $\bar{\nu}_e p \rightarrow \mu^+ n$).

Most of the model uncertainties are estimated by the GENIE reweighting infrastructure. From GENIE, the most significant sources of uncertainty comes from the normalization of the charge-current quasi-elastic cross section (1.60\%) and the axial mass parameter in the resonance cross section (1.41\%).

Uncertainties in the modification to the interaction model made by \minerva are relevant since the charged-current quasi-elastic (CCQE) scattering of electron-neutrinos is a significant source of background. The RPA correction uncertainties come from \cite{Gran:2017psn}, and the uncertainty on the event rate coming from this correction is 1.5\%. The uncertainty on the number of events coming from the tune to 2p2h interactions is estimated by comparing the effect on the simulation with and without the tune, and was found to be 1.52\%.  

This analysis tunes the normalization of the CCQE background using sidebands that capture events with high $Q^2_{ QE}$, and later the same normalization is applied to the low $Q^2_{ QE}$ background. A discrepancy in the shape of the CCQE background could lead to a incorrect estimate of the background in the signal region. In a similar way to Ref.\cite{PhysRevD.100.092001}, an analysis of \numubar CCQE-like events was used to compare the rate of events in the low and high transverse momentum at low recoil energy. It was found that although the simulation underestimates the data, it does by the similar amount in both $Q^2_{ QE}$ regions, and no further uncertainty is assigned to the shape of the CCQE Background.

The tree-level cross sections of neutrino-electron elastic scattering in GENIE are weighted to match  those which are calculated including radiative corrections \cite{radiativecorrection}. This updated cross section includes the production of real photons in the final state, and the fact that the energy measured in the detector is the sum of the final state electron and a photon. A 1.34\% systematic uncertainty is estimated by comparing the rate with and without this correction. A comparison of the corrected cross sections with GENIE is shown in appendix \ref{app:radcorrections}.

\begin{figure}[b]
    \centering
    \includegraphics[width=0.45\textwidth]{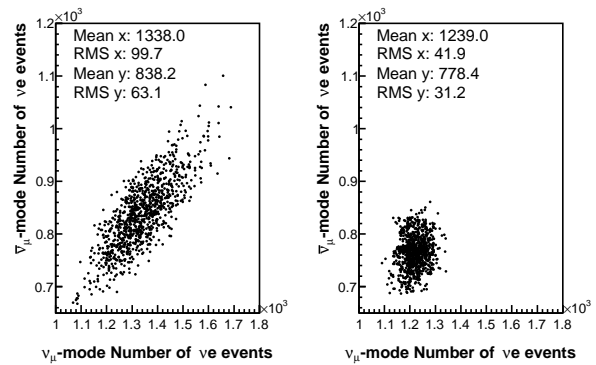}
    \caption{Predicted number of neutrino-electron elastic scattering events by each flux universe in \numu-mode and \numubar-mode. The left panel shows the distribution of the \textit{a priori} flux, the right panel shows the effect of the constraint on the predicted number of events. }
    \label{fig:neventsnue}
\end{figure}

\section{Flux Constraint}
\label{sec:flux}
Using the electron energy spectrum, it is possible to constrain the \numubar-mode flux with the same procedure as described in \cite{Park:2015eqa, PhysRevD.100.092001, DanIMD}. The measurement presented in this paper is used in combination with the other two ME constraints to get a single normalization constraint that can be applied to both \numu and \numubar beams.

The procedure is based on Bayes' theorem \cite{Tanabashi:2018oca}, in which the probability of a hypothesis given a measurement is proportional to the product of the \textit{a priori} probability of the hypothesis with the probability of the measurement given the hypothesis. In this case, the hypothesis is the neutrino flux prediction ($M$), and the measurement is the background-subtracted and efficiency corrected number of events measured at \minerva ($N$), such that
\begin{equation}
    \label{eq:bayes}
    P \left( M | N_{\nu e \rightarrow \nu e } \right) \propto P\left(M\right) P\left( N_{\nu e \rightarrow \nu e } | M \right).
\end{equation}
The \textit{a priori} flux uncertainty is estimated by using the multi-universe method \cite{Aliaga:2016oaz}. New predictions (universes) are created by randomly varying the underlying systematic parameters within their uncertainties while taking into account their correlations. Each flux universe yields a prediction for the number of electron elastic scattering events in \minerva. The flux uncertainty on the number of events is given by the spread of the universes. For each universe, the likelihood $P\left( N_{\nu e \rightarrow \nu e } | M \right)$ is calculated between the measured and predicted rate. The prediction from universes that have poor agreement with data are weighted down, reducing the spread of the universes leading to a lower flux uncertainty. The likelihood is \cite{Lyons86statisticsfor}:
\begin{equation}
\label{eq:fluxweight}
P\left( N_{\nu e \rightarrow \nu e } | M \right) = \frac{1}{(2\pi)^{K/2}}\frac{1}{|\Sigma_{\mathbf N}|^{1/2}} e^{-\frac{1}{2}\left({\mathbf N}-{\mathbf M}\right)^T\Sigma_{\mathbf N}^{-1}\left({\mathbf N}-{\mathbf M}\right)},
\end{equation}
where $\mathbf{N}$ is a vector of the content of the bins of the electron energy distribution. $\mathbf{M}$ is a vector of the bin content for the simulated prediction, $\Sigma_N$ is the covariance matrix of the measurements in $\mathbf{N}$ and $K$ is the number of bins on $\mathbf{N}$. For the $\nu e^{-}$ constraints, $\mathbf{N}$ and $\mathbf{M}$ contain the six bins of the electron energy distribution on either \numu-mode or \numubar-mode. For the combined $\nu e^{-}$ constraint the vectors contain a total of 12 bins using both beam modes. Finally, the combined $\nu e^{-}$+IMD includes the total number of inverse muon decay events measured on both modes in a single bin, yielding a total of 13 bins.

\begin{figure}[b]
    \centering
    \includegraphics[width=0.45\textwidth]{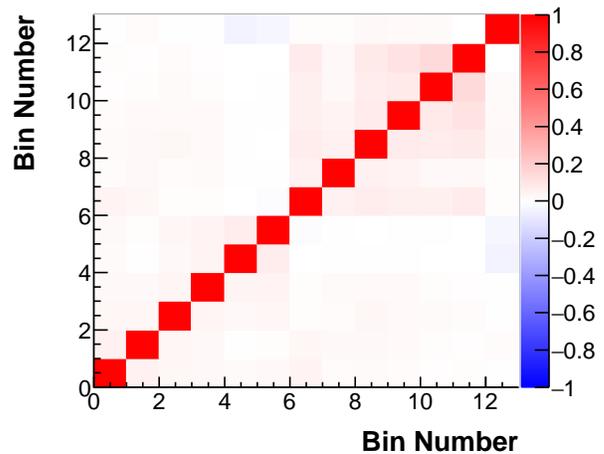}
    \caption{Correlation between the different measurements. Bin 0-5 are \numu-mode, 6-11 are \numubar-mode, 12 is total IMD events. The diagonal elements dominate since the leading uncertainty is statistics.}
    \label{fig:correlationmatrix}
\end{figure}

The flux uncertainty on the \numu and \numubar beams are correlated since the only differences on the NuMI beam configuration between the two modes are the polarity of the magnetic horns and the intensity of the beam. Additionally, because the hadron production constraints for $\pi^+$ and $\pi^-$ come from the same experiment, and because the largest uncertainties in those measurements are systematic and correlated between $\pi^+$ and $\pi^-$, a constraint on either charge meson will constrain production of both. It follows that the predictions for the number of events on each beam are correlated with one another as shown in the left panel of Fig. \ref{fig:neventsnue}. When combining the measurements, the covariance matrix of the systematic uncertainties is constructed. The systematic errors sources that are shared between the measurements are assumed 100$\%$ correlated. The correlation matrix is shown in Fig. \ref{fig:correlationmatrix} and the covariance matrix is tabulated on Table \ref{tab:covariance}. 

\begin{figure}[t]
\centering
  \includegraphics[width=0.45\textwidth]{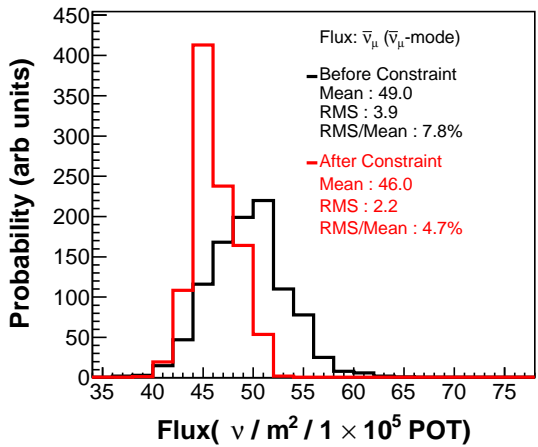}
\caption{Probability distributions of the \numubar in \numubar-mode flux between 2 and 20 GeV, before and after constraining the {\it a priori} flux model using the neutrino-electron scattering data.}
\label{fig:prob_dist_1}
\end{figure}

\begin{figure}[t]
\centering
  \includegraphics[width=0.45\textwidth]{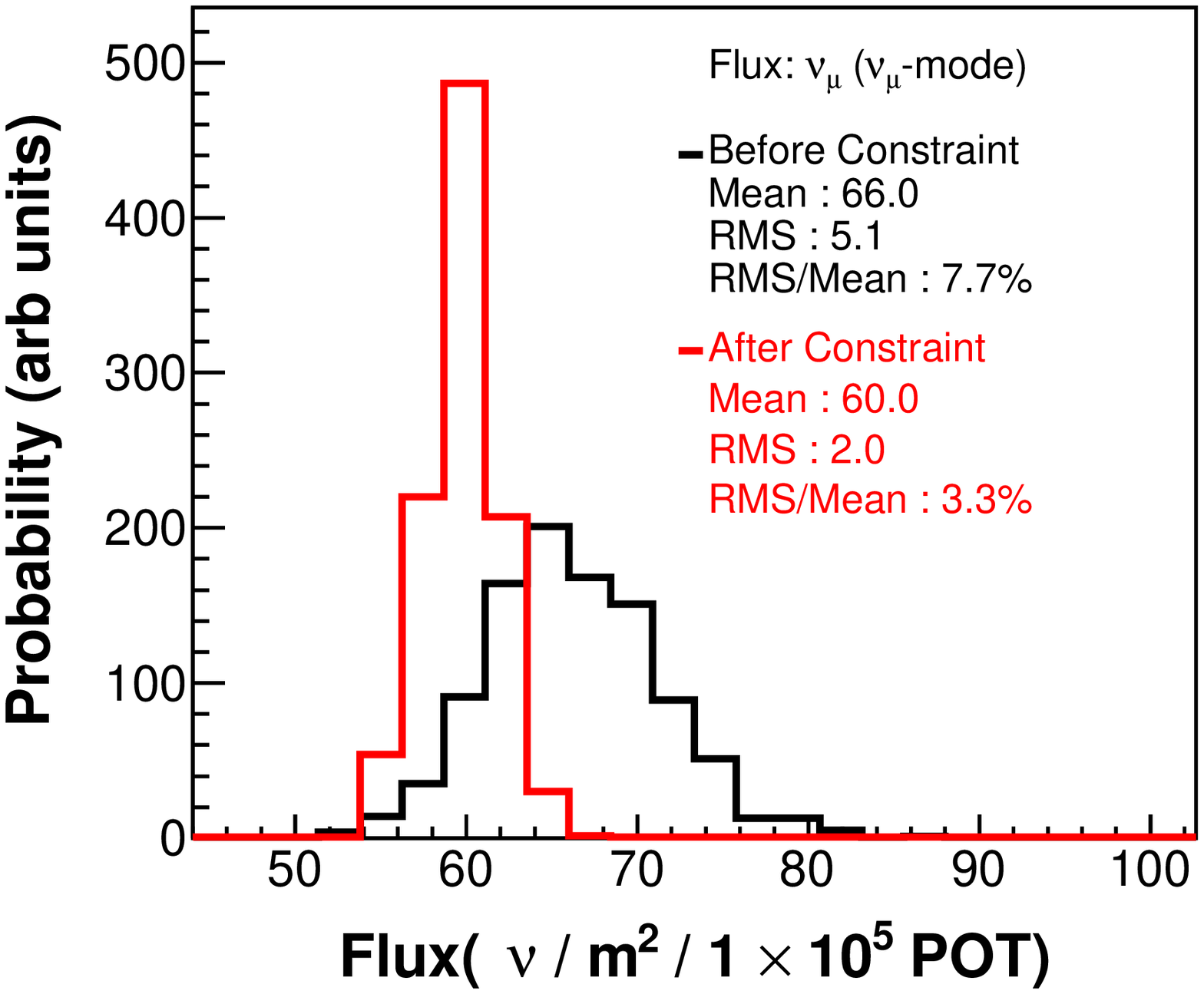}
\caption{Probability distributions of the \numu in \numu-mode flux between 2 and 20 GeV, before and after constraining the {\it a priori} flux model using the neutrino-electron scattering data.}
\label{fig:prob_dist_2}
\end{figure}

The results from applying the constraint using different measurements as inputs are shown in Table \ref{tab:fluxweights }. The three measurements independently are consistent with each other in the direction of the correction, lowering the neutrino flux predictions. The effect of the $\nu e$ constraints is stronger if used on the flux from which the measurement was made, and the greatest improvement on the constraint is achieved by combining both $\nu e^{-}$ measurements. The inverse muon decay measurement has a small effect, particularly on muon neutrinos on the \numu-mode and the wrong sign contamination components of each flux. Probability distributions for the predicted flux for \numubar-mode and \numu-mode of the NuMI beam are shown in Fig. \ref{fig:prob_dist_1} and Fig. \ref{fig:prob_dist_2}. To construct these plots, each flux universe is integrated between 2 and 20 GeV and then weighted by their respective likelihood according to Eq. \ref{eq:fluxweight}.

\begin{figure}
\centering
  \includegraphics[width=0.45\textwidth]{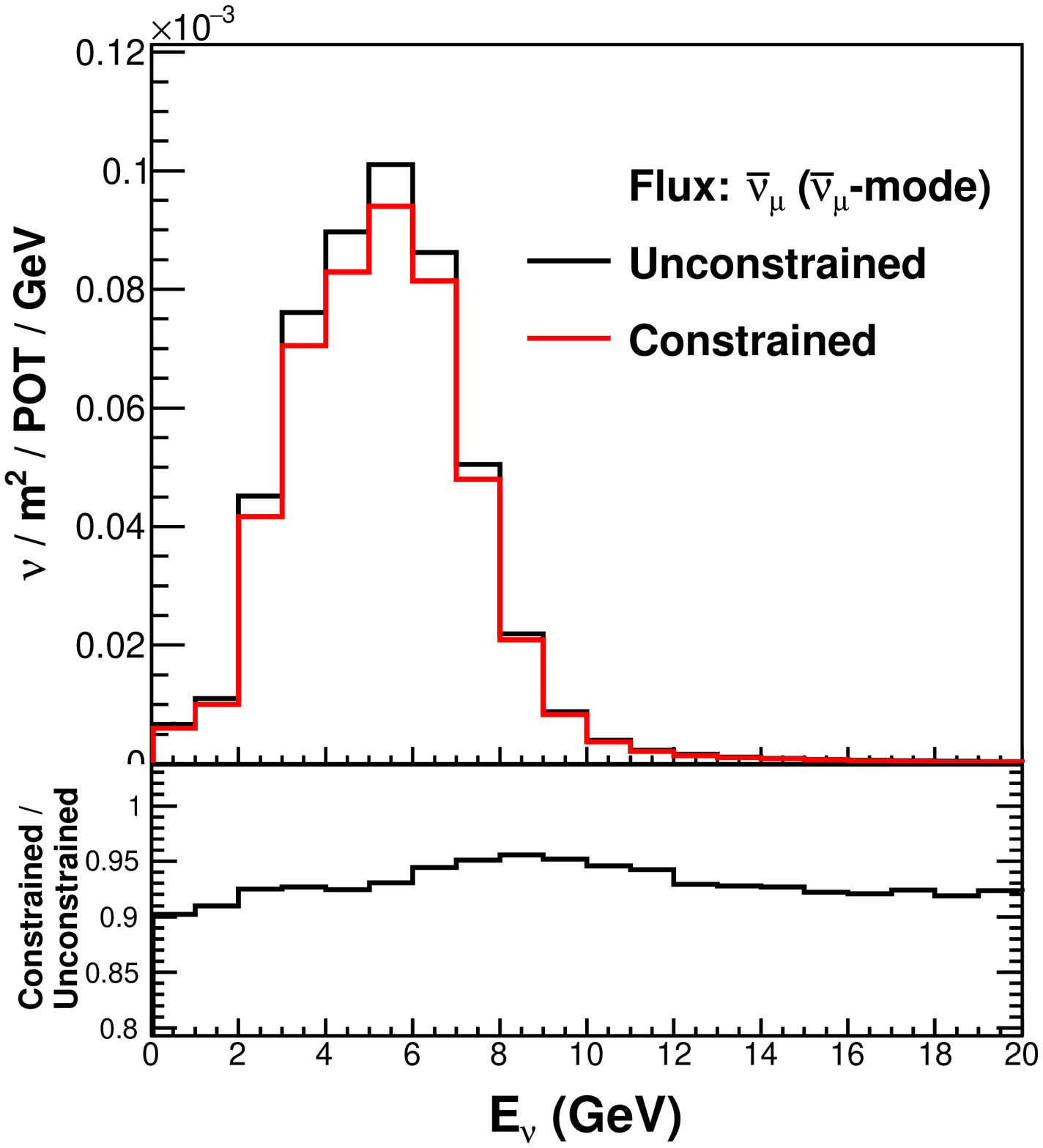} 
\caption{Predicted \numubar flux in bins of neutrino energy, before and after constraining the {\it a priori} flux model using the neutrino-electron scattering data.}
\label{fig:fluxconstraint1}
\end{figure}

\begin{figure}
\centering
  \includegraphics[width=0.45\textwidth]{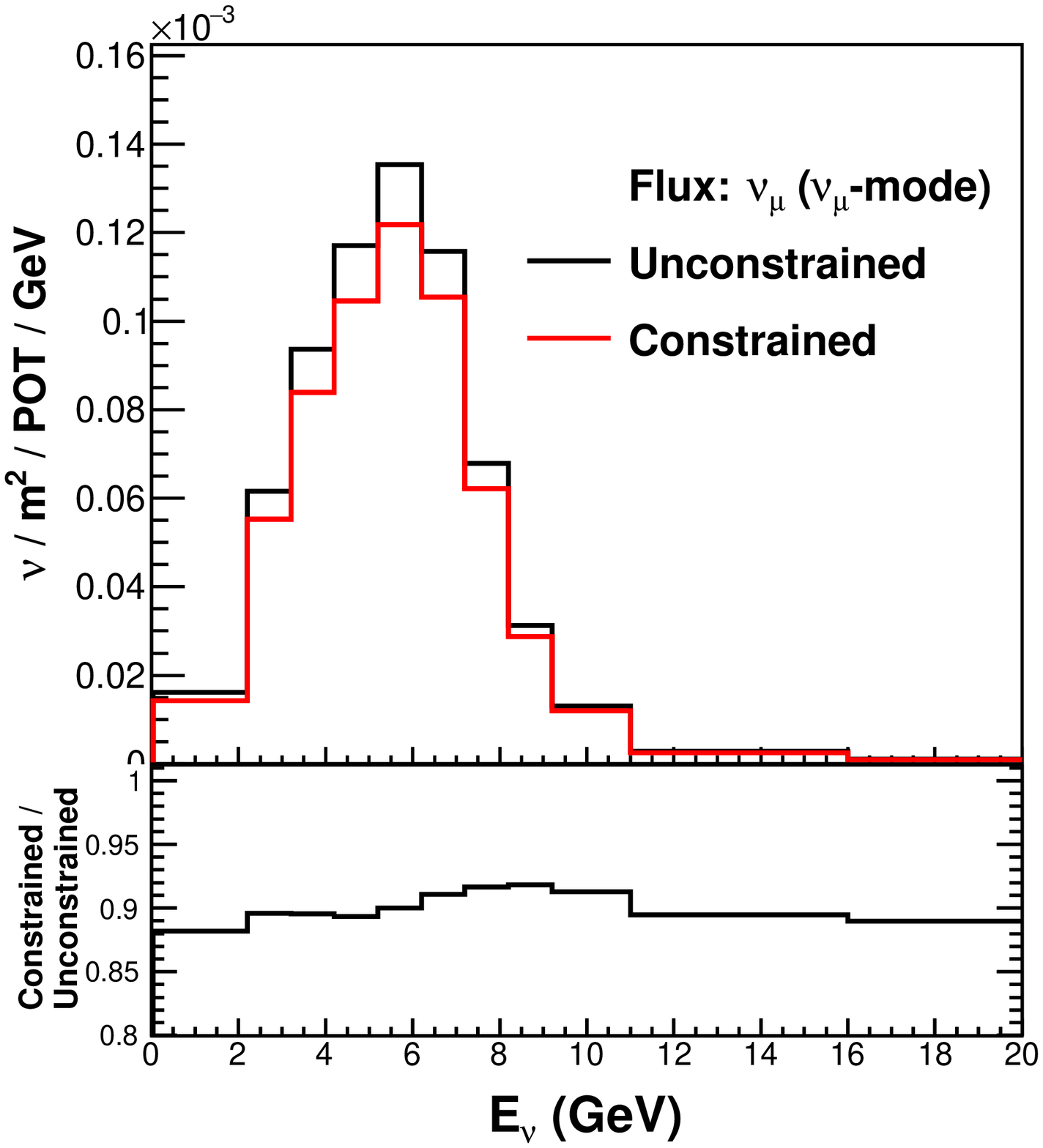} 
\caption{Predicted $\nu_\mu$ flux in bins of neutrino energy, before and after constraining the {\it a priori} flux model using the neutrino-electron scattering data.}
\label{fig:fluxconstraint2}
\end{figure}

\begin{figure}
\centering
  \includegraphics[width=0.45\textwidth]{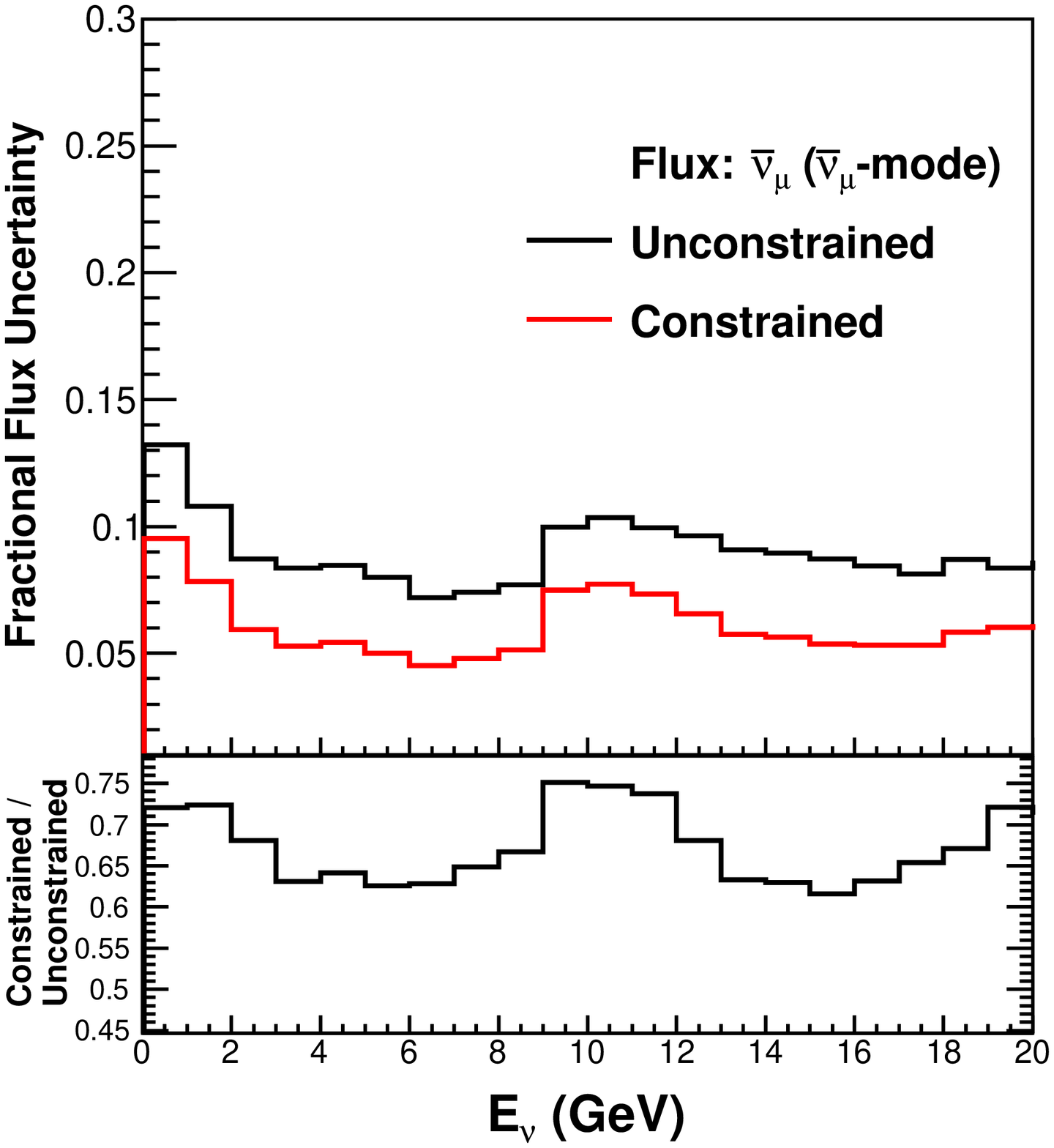}
\caption{Fractional uncertainties on the predicted \numubar flux in bins of neutrino energy, before and after constraining the {\it a priori} flux model using the neutrino-electron scattering data.}
\label{fig:fluxfractionaluncertainty1}
\end{figure}

\begin{figure}
\centering
  \includegraphics[width=0.45\textwidth]{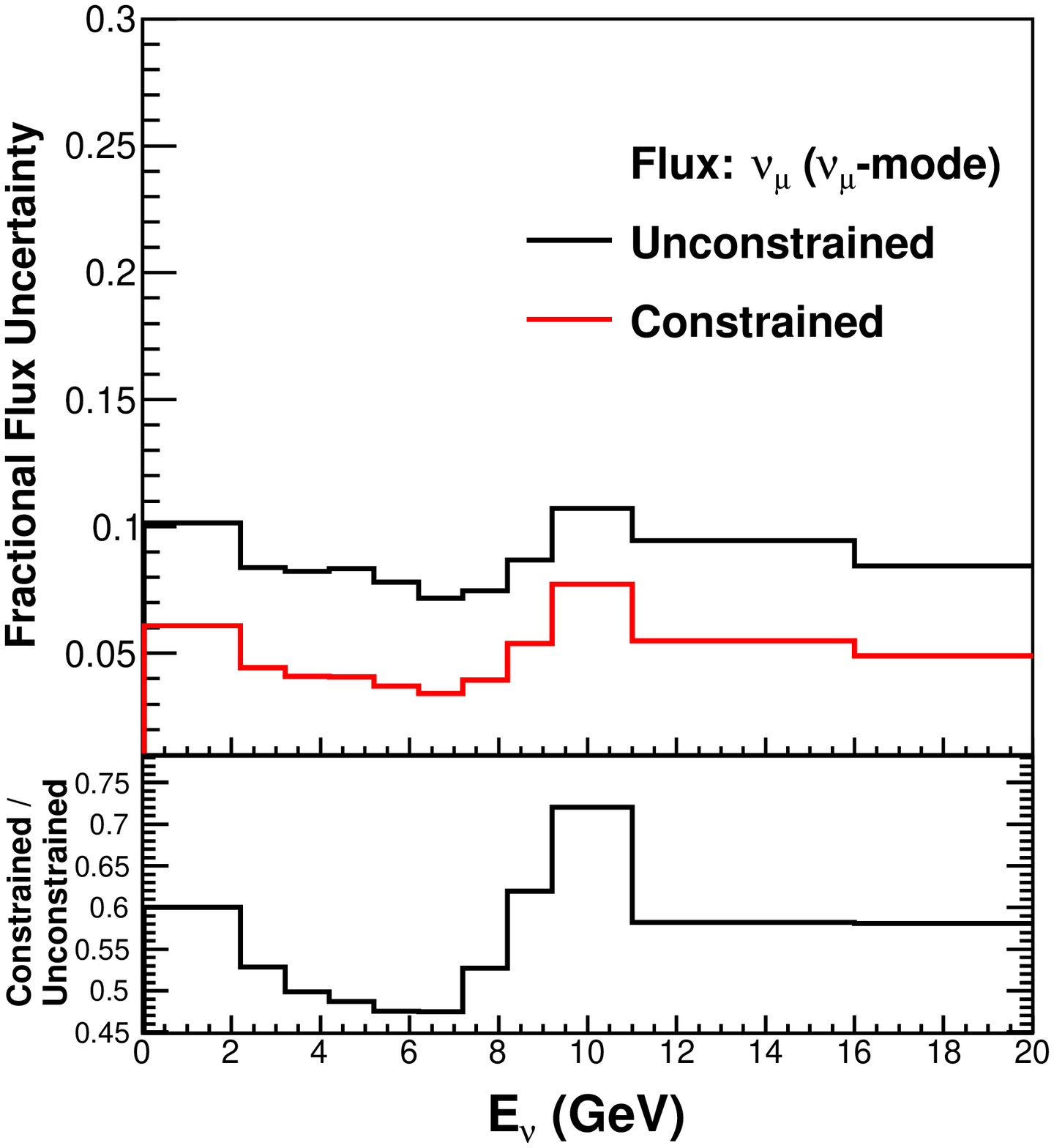}
\caption{Fractional uncertainties on the predicted $\nu_\mu$ flux in bins of neutrino energy, before and after constraining the {\it a priori} flux model using the neutrino-electron scattering data.}
\label{fig:fluxfractionaluncertainty2}
\end{figure}

The energy spectrum and fractional uncertainty on the neutrino flux before and after the constraint is shown in Figs. \ref{fig:fluxconstraint1}-\ref{fig:fluxfractionaluncertainty2}. The fractional uncertainties on the flux for the different neutrino species and different standard candle measurements are tabulated in table \ref{tab:fluxweights }. 

\begin{table*}
\centering
\caption{Estimated fractional systematic uncertainties (\%) of the neutrino flux for each flavor and polarity of the beam. First row show the uncertainty before constraint, the following rows represent the measurement from which the constraint was calculated.}
\begin{tabular}{l|cccc|cccc}
\hline
\hline
& \multicolumn{4}{c}{\numubar-mode} & \multicolumn{4}{c}{\numu-mode} \\
& \numubar & $\bar{\nu}_e$  & \numu &  $\nu_e$  & \numu  & $\nu_e$  & \numubar &  $\bar{\nu}_e$  \\
\hline
\textit{a priori} &      7.76 & 7.81 & 11.1 & 11.9 & 7.62 & 7.52 & 12.2 & 11.7 \\
\numu-mode $\nu e^{-}$ &     6.11 & 5.81 & 6.30 & 8.50 & 3.90 & 3.94 & 8.37 & 8.68 \\
\numubar-mode $\nu e^{-}$ &  4.92 & 4.98 & 8.07 & 9.19 & 5.88 & 5.68 & 8.36 & 8.64 \\
combined $\nu e^{-}$ &       4.68 & 4.62 & 5.56 & 7.80 & 3.56 & 3.58 & 7.15 & 7.84 \\
combined $\nu e^{-}$ + IMD & 4.66 & 4.56 & 5.20 & 6.08 & 3.27 & 3.22 & 6.98 & 7.54 \\
\hline
\hline
\end{tabular}
\label{tab:fluxweights }
\end{table*}


\section{Conclusion}
\label{sec:conclusion}
This article presents the electron energy spectrum for a sample of antineutrino-electron elastic scattering events observed in the \minerva detector during the NuMI ME \numubar-mode run. A total of 578 events were observed after background subtraction, and corresponds to an exposure of 1.12 $\times$ 10$^{21}$ protons on target. When this sample is combined with the \numu-mode \cite{PhysRevD.100.092001} and the IMD \cite{DanIMD} results, the uncertainty on the \numu (\numubar) flux during the \numu (\numubar) mode operation has been reduced from 7.6$\%$ (7.8$\%$) to 3.3$\%$ (4.7$\%$).
The improved flux prediction will benefit future \minerva cross section measurements that use the ME beam. This technique can also be used by future neutrino oscillation experiments such as DUNE \cite{Marshall:2019vdy}.

\begin{acknowledgments}
This document was prepared by members of the MINERvA Collaboration using the resources of the Fermi National Accelerator Laboratory (Fermilab), a U.S. Department of Energy, Office of Science, HEP User Facility. Fermilab is managed by Fermi Research Alliance, LLC (FRA), acting under Contract No. DE-AC02-07CH11359. 
These resources included support for the \minerva construction project, and support for construction also was granted by the United States National Science Foundation under Award No. PHY-0619727 and by the University of Rochester. Support for participating scientists was provided by NSF and DOE (USA); by CAPES and CNPq (Brazil); by CoNaCyT (Mexico); by Proyecto Basal FB 0821, CONICYT PIA ACT1413, and Fondecyt 3170845 and 11130133 (Chile); by CONCYTEC (Consejo Nacional de Ciencia, Tecnolog\'ia e Innovaci\'on Tecnol\'ogica), DGI-PUCP (Direcci\'on de Gesti\'on de la Investigaci\'on - Pontificia Universidad Cat\'olica del Peru), and VRI-UNI (Vice-Rectorate for Research of National University of Engineering) (Peru); NCN Opus Grant No. 2016/21/B/ST2/01092 (Poland); by Science and Technology Facilities Council (UK); by EU Horizon 2020 Marie Sk lodowska-Curie Action; by a Cottrell Postdoctoral Fellowship from the Research Corporation for Scientific Advancement; by an Imperial College London President’s PhD Scholarship. 
We thank the MINOS Collaboration for use of its near detector data. Finally, we thank the staff of Fermilab for support of the beam line, the detector, and computing infrastructure.

\end{acknowledgments}

\appendix
\section{Radiative corrections to the GENIE Neutrino-Electron Scattering Model}
\label{app:radcorrections}
\begin{figure}
    \centering
    \includegraphics[width=0.45\textwidth]{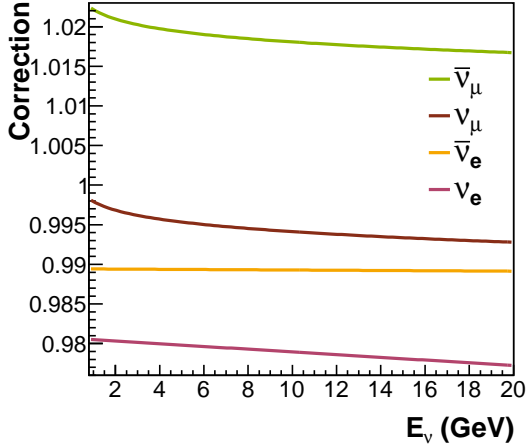}
    \caption{The ratio between the the cross section predicted by \cite{radiativecorrection} and GENIE neutrino-electron elastic scattering cross sections is used as a correction on the simulation as a function of true neutrino energy.}
    \label{fig:raditaticecorrection}
\end{figure}

At leading order, the neutrino-electron scattering cross section is given by 
\begin{align}
 &\frac{d\sigma(\nu e^-\to\nu e^-)}{dy} \\ 
 & = \frac{G^2_F s}{\pi}\left[C_{LL}^2+C_{LR}^2(1-y)^2-C_{LL}C_{LR}\frac{m y}{E_\nu}\right] ,
\label{eqn:tree-xsec}
\end{align}
where $E_\nu$ is the neutrino energy, $s$ is the Mandelstam
invariant representing the square of the total energy in the center-of-mass frame, $m$ is the electron mass, and $y=T_e/E_\nu$, where $T_e$ is the kinetic energy of the final state electron. The expression for the related antineutrino process can be obtained by interchanging the couplings $C_{LL}$ and $C_{LR}$ in Eq. \ref{eqn:tree-xsec}. In GENIE 2.12.6 the couplings are $C_{LL}^{\nu_\mu e}=\text{-0.2723}$, $C_{LL}^{\nu_e e}=\text{0.7277}$ and $ C_{LR}=\text{0.2277}$.
The next-to-leading order radiative corrections, which include contribution of a real photon in the final state, have been calculated by\cite{radiativecorrection}. A ratio is taken between the absolute cross section predictions in Ref.\cite{radiativecorrection} and GENIE 2.12.6 neutrino-electron elastic scattering cross section. The ratio is applied as a weight to the simulated neutrino electron elastic scattering event as a function of true neutrino energy. Fig. \ref{fig:raditaticecorrection} shows the correction applied to the different neutrino flux components. 

\begin{table*}
\centering 
\caption{Covariance matrix of the measurements used in the calculation of the combined constrain. The bin range of each bin is shown for the electron elastic scattering results. The error for the inverse muon decay is the error on the total number of events. The covariance from the \numu-mode is from \cite{PhysRevD.100.092001}, and for IMD it is from the results of \cite{DanIMD}. The covariance from \numubar-mode and that between the different measurements is a result of the analysis.}
\begin{tabular}{ l c | *{6}{r} | *{6}{r} | r}
\hline\hline
& &\multicolumn{6}{c}{\numu-mode $\nu e^{-}$} & \multicolumn{6}{c}{\numubar-mode $\nu e^{-}$} & IMD  \\
\hline
\multicolumn{2}{c|}{Bin Range (GeV)}  &  0.8-2 &   2-3   &  3-5   & 5-7    & 7-9    & 9-$\infty$ & 0.8-2 &   2-3   &  3-5   & 5-7    & 7-9    & 9-$\infty$ & N/A \\
\multicolumn{2}{c|}{Bin content} & 329.68 & 200.88 & 310.05 & 167.62 & 78.77 & 101.47 & 218.03 & 133.26 & 247.64 & 97.12 & 47.66 & 54.65 & 183.21 \\
\hline
\multirow{6}{*}{\numu-mode $\nu e^{-}$}  & 0.8-2 & 938.15 &  31.55 &  27.73 &  16.00 &   8.22 &  17.51 &  36.44 &   8.03 &  15.54 &   8.15 &   3.72 &   7.43 &  -2.76 \\ 
                                         & 2-3   &  31.55 & 376.23 &  16.91 &   9.71 &   1.46 &   5.60 &  18.95 &   8.82 &  12.03 &   8.71 &   3.35 &   3.53 &   8.29 \\ 
                                         & 3-5   &  27.73 &  16.91 & 558.98 &  16.84 &   9.02 &  17.25 &   7.05 &   7.36 &  18.90 &  10.30 &   5.87 &   6.78 &  -2.44 \\ 
                                         & 5-7   & 16.00 &   9.71 & 16.84 & 324.21 &   9.62 &  16.88 &   5.17 &   6.05 &  10.49 &   7.92 &   3.18 &   2.36 &  -2.62 \\
                                         & 7-9   &  8.22 &   1.46 & 9.02 &   9.62 & 162.75 &  18.19 &   0.01 &   1.19 &   1.20 &   1.31 &  -0.16 &  -0.64 & -14.07 \\ 
                                         & 9-$\infty$ & 17.51 &   5.60 &  17.25 &  16.88 &  18.19 & 384.95 &  -5.77 &   1.22 &   0.49 &   1.87 &   0.73 &  -0.65 & -16.46 \\
\hline
\multirow{6}{*}{\numubar-mode $\nu e^{-}$}   &0.8-2 &  36.44 &  18.95 &   7.05 &   5.17 &   0.01 &  -5.77 & 674.31 &  23.44 &  41.42 &  26.11 &  18.57 &  37.42 &   8.04 \\ 
                                             & 2-3  &   8.03 &   8.82 &   7.36 &   6.05 &   1.19 &   1.22 &  23.44 & 270.59 &  20.62 &  13.01 &   5.68 &   9.07 &   4.98 \\ 
                                             &3-5  &  15.54 &  12.03 &  18.90 &  10.49 &   1.20 &   0.49 &  41.42 &  20.62 & 500.10 &  30.35 &  20.93 &  33.27 &  14.92 \\ 
                                             &5-7  &   8.15 &   8.71 &  10.30 &   7.92 &   1.31 &   1.87 &  26.11 &  13.01 &  30.35 & 284.50 &  18.46 &  33.50 &   9.93 \\ 
                                             &7-9  &   3.72 &   3.35 &   5.87 &   3.18 &  -0.16 &   0.73 &  18.57 &   5.68 &  20.93 &  18.46 & 151.92 &  31.64 &   5.61 \\ 
                                             &9-$\infty$  &  7.43 &   3.53 &   6.78 &   2.36 &  -0.64 &  -0.65 &  37.42 &   9.07 &  33.27 &  33.50 &  31.64 & 321.20 &  -1.07 \\ 
\hline
IMD                                          & N/A   & -2.76 &   8.29 &  -2.44 &  -2.62 & -14.07 & -16.46 &   8.04 &   4.98 &  14.92 &   9.93 &   5.61 &  -1.07 & 552.72 \\ 
\end{tabular}
\label{tab:covariance}
\end{table*}

\bibliographystyle{apsrev4-2}
\bibliography{fluxconstraint}

\end{document}